\documentclass[conference]{IEEEtran}

\usepackage{amsmath,amssymb,amsfonts}
\usepackage{algorithmic}
\usepackage{algorithm}
\usepackage{graphicx}
\usepackage{textcomp}
\usepackage{xcolor}
\usepackage{booktabs}
\usepackage{tabularx}
\usepackage{multirow}
\usepackage{cite}
\usepackage{url}
\usepackage{bm}
\usepackage{hyperref}

\begin{document}

\title{Protein folding on a 64 qubit trapped-ion hardware via counterdiabatic quantum optimization}

\author{
\IEEEauthorblockN{Alejandro Gomez Cadavid$^{1,2}$, 
Pavle Nika\v{c}evi\'{c}$^{1}$, 
Pranav Chandarana$^{1,2}$,
Sebastián V. Romero$^{3,4}$,\\
Enrique Solano$^{1}$,
Narendra N. Hegade$^{1,5}$
}\medskip
\IEEEauthorblockA{$^{1}$Kipu Quantum GmbH, Greifswalderstrasse 212, 10405 Berlin, Germany\\
$^{2}$Department of Physical Chemistry, University of the Basque Country EHU, Apartado 644, 48080 Bilbao, Spain\\
$^{3}$Instituto de Ciencia de Materiales de Madrid (ICMM-CSIC), Cantoblanco, E-28049 Madrid, Spain\\
$^{4}$Departamento de Física Teórica de la Materia Condensada, Universidad Autónoma de Madrid, E-28049 Madrid, Spain\\
$^{5}$IDAL, Electronic Engineering Department, ETSE-UV, University of Valencia,\\ Avgda. Universitat s/n, 46100 Burjassot, Valencia, Spain}\\
\IEEEauthorblockN{
Miguel Angel Lopez-Ruiz$^{6}$,
Claudio Girotto$^{6}$,
Hanna Linn$^{6}$,
Hakan Doga$^{6}$,
Evgeny Epifanovsky$^{6}$,\\
Panagiotis Kl. Barkoutsos$^{6}$,
Ananth Kaushik$^{6}$,
Martin Roetteler$^{6}$
}\medskip
\IEEEauthorblockA{
$^{6}$IonQ Inc., 4505 Campus Dr, College Park, MD 20740, USA}
}

\maketitle

\begin{abstract}
We report the largest trapped-ion hardware demonstration of lattice protein-folding optimization to date, using bias-field digitized counterdiabatic quantum optimization (BF-DCQO) on a fully connected 64-qubit Barium development system similar to the forthcoming IonQ Tempo line. Six peptide sequences with 14--16 amino-acid residues are encoded using a coarse-grained tetrahedral lattice model, yielding higher-order spin-glass Hamiltonians with long-range interactions involving up to five-body terms and mapped to 46--61 qubits. The resulting instances are demanding for near-term quantum hardware because low-energy configurations must satisfy backbone-geometry constraints while optimizing dense residue-contact interactions. BF-DCQO uses a non-variational bias-feedback mechanism, where low-energy samples from each round define longitudinal fields that guide subsequent quantum evolutions. Across the studied instances, BF-DCQO shifts raw sampled energy distributions toward lower energies than uniform random sampling, with the strongest improvements appearing in residue-contact variables. To preserve this signal, we introduce a consensus-based post-processing pipeline that combines quantum-learned contact information with feasible backbone geometries. The resulting hybrid workflow reaches the classical reference energy in multiple instances and improves over the corresponding random-seeded pipeline. These results show that BF-DCQO can generate structured samples for dense protein-folding Hamiltonians at previously unexplored trapped-ion scales.
\end{abstract}

\section{Introduction}

Predicting the three-dimensional structure of a protein from its amino acid sequence, known as the protein folding problem, remains one of the grand challenges in computational biology~\cite{dill2012,anfinsen1973}. The configurational space grows exponentially with chain length, and finding the minimum-energy conformation is NP-hard even on simplified lattice models. While deep-learning approaches have achieved remarkable accuracy for known protein families, the underlying combinatorial optimization naturally maps to formulations amenable to quantum computation~\cite{perdomo2012,robert2021}.

Lattice models reduce the continuous folding problem to a discrete optimization over turn sequences on a regular lattice, producing Ising Hamiltonians that quantum hardware can target directly. Reference~\cite{perdomo2012} demonstrated quantum annealing on small peptides using D-Wave hardware, while Ref.~\cite{babej2018} explored coarse-grained models on quantum annealers. Reference~\cite{robert2021} introduced a tetrahedral lattice encoding with physically realistic Miyazawa-Jernigan interaction potentials, capturing some important features of protein geometry while producing gate-model-compatible Ising Hamiltonians. Subsequent research branched in several directions. Algorithmic alternatives, such as digitized-counterdiabatic methods~\cite{chandarana2022} and later BF-DCQO, pushed hardware demonstrations up to 12 amino acids using 33 trapped-ion qubits~\cite{romero2025}. Critical benchmarking and resource studies showed both the limitations of current quantum hardware on simplified peptide-folding tasks and the costs of existing encodings~\cite{Boulebnane2023, linn2024}. Finally, representational extensions were introduced, including alternative turn-based encodings that reached 20 amino acids using 114 superconducting qubits~\cite{vasavi2024}; higher-degree-of-freedom cubic and FCC lattices~\cite{li2025}; as well as problem-agnostic approaches, which reached up to 26 amino acids using 45 superconducting qubits~\cite{linn2025}. 

Variational algorithms such as the Quantum Approximate Optimization Algorithm (QAOA)~\cite{farhi2014} and the Variational Quantum Eigensolver (VQE)~\cite{peruzzo2014} have been the primary gate-model approaches for combinatorial optimization, but they require iterative classical-quantum feedback loops that are difficult to scale. Digitized Counterdiabatic Quantum Optimization (DCQO)~\cite{hegade2022,chandarana2022} offers an alternative: it combines counterdiabatic (CD) driving~\cite{sels2017,claeys2019} with digital quantum circuits, using analytically derived circuit parameters with no variational optimization. For short evolution times, the dynamics are dominated by the CD term, yielding shallow circuits. The bias-field extension (BF-DCQO)~\cite{cadavid2024, romero2024} adds an iterative refinement, where low-energy measurement outcomes feed back as longitudinal bias fields in subsequent rounds, progressively increasing initial state's overlap with low-energy configurations. The bias update is a simple closed-form operation (later given by Eq.~\ref{eq:bias_update}) that avoids gradient estimations.

BF-DCQO has demonstrated strong performance on higher order binary optimization problems~\cite{romero2024}. Here, we apply it to protein folding on trapped-ion hardware beyond the previously achieved scale (33 qubits~\cite{romero2025}), tackling instances up to 16 amino acids using 61 all-to-all connected qubits on a 64-qubit Barium development system similar to the forthcoming IonQ Tempo line. Trapped-ion platforms are particularly well-suited for this problem, as their all-to-all connectivity eliminates the SWAP overhead required by nearest-neighbor architectures to implement the dense Ising couplings arising from pairwise bead interactions~\cite{linke2017experimental}.
Therefore, to the best of our knowledge, this work contains the largest protein-folding demonstration on trapped-ion hardware up to date, as well as the largest one leveraging the dense encoding originally proposed in Ref.~\cite{robert2021}.

\textbf{Our contributions:} (1)~We present the first application of BF-DCQO to protein folding in the range of 46--61 qubits on trapped-ion hardware. (2)~We demonstrate that BF-DCQO captures meaningful interaction structure, reaching the exact reference energy in 4 of the 6 studied sequences when paired with a consensus-based post-processing. How to choose the post-processing is a central question for any demonstration on current quantum hardware. Then, we also show that the consensus pipeline can preserve the quantum-learned contacts, while a more naive per-sample repair, which only re-optimizes contact qubits, can erase the structure learned by the quantum algorithm. (3)~We analyze the structure of the quantum signal, finding that it concentrates in the interaction qubits, consistent with how the bias-field mechanism operates.

\section{Problem Formulation}

\subsection{Tetrahedral Lattice Model}

We adopt the coarse-grained lattice protein model of Ref.~\cite{robert2021}, ignoring side-chains. Each amino acid residue is represented by a main-chain bead placed on the vertices of a tetrahedral lattice. The protein backbone is encoded as a self-avoiding walk, in which each turn between consecutive beads is one of 4 tetrahedral directions, requiring 2 qubits per turn. The total Hamiltonian is the sum of two terms,

\begin{equation}\label{eq:hamiltonian}
  H = H_{\text{back}} + H_{\text{contact}},
\end{equation}

where $H_{\text{back}}$ ensures physical geometries by penalizing consecutive turns along the same axis (backtracking), and $H_{\text{contact}}$ encodes pairwise contact interactions between backbone beads that are separated by five or more residues along the chain but come into spatial proximity due to folding, which induces an interaction defined by the Miyazawa--Jernigan potential~\cite{miyazawa1985} while penalizing overlapping conformations to enforce the self-avoiding walk. The constraint term $H_{\text{back}}$ carries large penalty coefficients to ensure that the ground state of~$H$ corresponds to a physically valid conformation. The interaction contact terms ($H_{\text{contact}}$) encode the physical energetics of self-avoiding sequences, where lower contact energy indicates a more favorable folded structure.

\subsection{Qubit Layout}

The total number of qubits is $n_\text{q} = n_{\text{contact}} + n_{\text{geom}}$:
\begin{itemize}
\item \textbf{Geometry qubits} ($n_{\text{geom}}$): encode the turn sequence on the tetrahedral lattice. These must respect the no-backtracking constraints.
\item \textbf{Contact qubits} ($n_{\text{contact}}$): indicator variables for pairwise bead interactions. For a fixed backbone geometry, the contact configuration can be determined efficiently by inspecting the 3D structure.
\end{itemize}

Our instances range from 46 to 61 qubits, see Table~\ref{tab:sequences}. Specifically, 14-residue chains yield 46 qubits ($n_{\text{contact}} = 25$, $n_{\text{geom}} = 21$), $15$-residue chains yield 53 qubits ($n_{\text{contact}} = 30$, $n_{\text{geom}} = 23$), and $16$-residue chains yield 61 qubits ($n_{\text{contact}} = 36$, $n_{\text{geom}} = 25$). In this qubit layout, the contact qubits constitute the majority of the total qubit count. The chosen sequences are presented in Table~\ref{tab:sequences}. Particularly, IDWKKLLDAAKQIL~\cite{IDWKKLLDAAKQIL} is mastoparan~I, a venom-derived peptide from \textit{Polybia paulista} with antimicrobial and immunomodulatory activity. RGKWTYNGITYEGR~\cite{RGKWTYNGITYEGR} is a \textit{de novo} designed $\beta$-hairpin-forming peptide identified via combinatorial screening, belonging to a class of short peptides widely used as model systems for investigating $\beta$-sheet folding and stability. KWKLFKKIGAVLKVL~\cite{KWKLFKKIGAVLKVL} is CM15, a synthetic hybrid of cecropin~A and melittin with a highly alpha-helical structure and a broad-spectrum antimicrobial activity via membrane disruption. LEPFSGKALCSWSIC~\cite{LEPFSGKALCSWSIC} is a fragment of tissue transglutaminase (TGM2), an enzyme involved in extracellular matrix stabilization, apoptosis regulation, and wound healing. MRWQEMGYIFYPRKLR~\cite{MRWQEMGYIFYPRKLR} corresponds to MOTS-c, a mitochondrial-derived peptide that regulates metabolism through the regulator AMPK activation and insulin sensitization.
VARGWKRKCPLFGKGG~\cite{VARGWKRKCPLFGKGG} corresponds to VG16KRKP, a cationic antimicrobial peptide that neutralizes bacterial endotoxins via electrostatic interactions with lipopolysaccharides.

\begin{table}[htb]
\centering
\caption{Sequences used in this work as a benchmark.}
\label{tab:sequences}
\begin{tabular}{lcccc}
\toprule
Sequence & Length & $n_\text{q}$ & $n_\text{geom}$ & $n_\text{contact}$ \\
\midrule
IDWKKLLDAAKQIL   & $14$ & $46$ & $21$ & $25$ \\
RGKWTYNGITYEGR   & $14$ & $46$ & $21$ & $25$ \\
KWKLFKKIGAVLKVL  & $15$ & $53$ & $23$ & $30$ \\
LEPFSGKALCSWSIC  & $15$ & $53$ & $23$ & $30$ \\
MRWQEMGYIFYPRKLR & $16$ & $61$ & $25$ & $36$ \\
VARGWKRKCPLFGKGG & $16$ & $61$ & $25$ & $36$ \\
\bottomrule
\end{tabular}
\end{table}

\section{Methods}

\subsection{BF-DCQO Algorithm}

\emph{Counterdiabatic Driving.--} Consider the adiabatic Hamiltonian interpolating between an initial Hamiltonian $H_\text{i}$ and the problem Hamiltonian $H_\text{f}$, taken to be the one from Eq.~\ref{eq:hamiltonian}

\begin{equation}\label{eq:adiabatic}
H_{\text{ad}}(\lambda) = (1-\lambda)\,H_i + \lambda\,H_f, \qquad \lambda \in [0,1].
\end{equation}
The exact counterdiabatic term $\mathcal{A}_\lambda$ suppresses diabatic transitions during a finite-time evolution. We approximate it via the first-order nested commutator ansatz~\cite{sels2017,claeys2019}
\begin{equation}\label{eq:cd}
\mathcal{A}_\lambda \approx i\,\alpha_1(\lambda)\,[H_i, H_f],
\end{equation}
where 
\begin{equation}
\alpha_1(\lambda) = - \frac{||[H_i, H_f]||_F^2}{ (1-\lambda) ||[H_i, [H_i, H_f]]||_F^2 + \lambda ||[H_f, [H_i, H_f]]||_F^2 }    
\end{equation}

\emph{Impulse Regime.--} For short total evolution time $T$, the system dynamics are dominated by $\mathcal{A}_\lambda$, yielding the \emph{impulse} unitary~\cite{hegade2022}:
\begin{equation}
U \approx \exp\!\bigl(-i\,\Delta t\;\mathcal{A}_\lambda\bigr).
\end{equation}
This is digitized via Trotterization over $n_{\text{steps}}$ layers. To reduce circuit depth, Pauli terms with coefficients $|\Delta t \cdot r_{jk}| < \theta$ are pruned, where $\theta$ is a cutoff threshold. More aggressive pruning yields shallower circuits at the cost of approximation fidelity.

\emph{Bias-Field Updates.--} The standard driver $H_i = -\sum_j X_j$ is generalized to include bias fields:
\begin{equation}\label{eq:driver}
H_i^{(r)} = \sum_j \bigl(-X_j + h_j^{(r)}\,Z_j\bigr),
\end{equation}
where $h_j^{(r)}$ is the bias field for qubit $j$ at iteration $r$. After each measurement round, the bias fields are updated from the low-energy subset $S_l$:
\begin{equation}\label{eq:bias_update}
h_j^{(r+1)} = -K_s \langle \sigma_j^z \rangle_{S_l},
\end{equation}
where $S_l$ contains the $n_l$ lowest-energy bitstrings measured in round $r$ and $K_s$ is a scaling factor that controls the strength of the bias update. This drives the initial state toward regions of the Hilbert space populated by good solutions. The full procedure is summarized in Algorithms~\ref{alg:dcqo} and~\ref{alg:bfdcqo}.

\begin{algorithm}[t]
\caption{DCQO~\cite{cadaviddcqo2024}}\label{alg:dcqo}
\begin{algorithmic}[1]
\REQUIRE Target Hamiltonian $H_f$, initial Hamiltonian $H_i$, total evolution time $T$, number of steps $n_s$, threshold $\theta$
\STATE Set $\Delta t \gets T / n_s$
\STATE Construct $\mathcal{A}_\lambda = i\alpha_1[H_i, H_f]$
\STATE Discard contributions with $|\Delta t \cdot r_{jk}| < \theta$
\FOR{$l = 1$ to $n_s$}
  \STATE Implement Trotter step: $U_l = \exp(-i\,\Delta t\,\mathcal{A}_\lambda)$
\ENDFOR
\STATE Perform measurements in the computational basis
\RETURN sampled bitstring $\bm{z}$
\end{algorithmic}
\end{algorithm}

\begin{algorithm}[t]
\caption{BF-DCQO~\cite{cadavid2024, romero2024}}\label{alg:bfdcqo}
\begin{algorithmic}[1]
\REQUIRE $H_f$, number of rounds $R$, number of shots $n_{\text{shots}}$, number of selected states $n_l$, parameters $T$, $n_s$, $\theta$
\STATE Initialize local fields $h_j^{(1)} \gets 0$ for all qubits $j$
\FOR{$r = 1$ to $R$}
  \STATE Define $H_i^{(r)} = \sum_j (-X_j + h_j^{(r)} Z_j)$
  \STATE Execute DCQO$(H_f, H_i^{(r)}, T, n_s, \theta)$ using $n_{\text{shots}}$ samples
  \STATE Compute energies $E(\bm{z})$ for all observed bitstrings $\bm{z}$
  \STATE Let $S_l$ be the subset of $n_l$ bitstrings with lowest energies
  \STATE Update fields: $h_j^{(r+1)} \gets - K_s \langle \sigma_j^z \rangle_{S_l}$ for each qubit $j$
\ENDFOR
\RETURN collection of all sampled bitstrings from every round
\end{algorithmic}
\end{algorithm}

We use the following parameters for the BF-DCQO runs: $R = 10$ iterations ($8$ for MRWQEMGYIFYPRKLR, defined upon hardware availability), $n_{\text{shots}} = 5,000$ per iteration, $T = 1$, $n_{\text{steps}} = 1$, $K_s=2$, $n_l = 100$ low-energy samples for bias updates. Two pruning levels are tested, high (aggressive pruning, shallower circuits) and low (conservative, deeper circuits). For each configuration, an equal number of uniform random bitstrings serves as a point of comparison.

\subsection{Experimental Setup}

Experiments were executed on a 64-qubit Barium development system similar to the forthcoming IonQ Tempo line~\cite{chen2024,delaney2024} with up to 64 Barium qubits, a long-chain steered-beam architecture with 532\,nm laser pulses and all-to-all qubit connectivity via collective motional modes enabling arbitrary single-qubit
rotations and entangling ZZ gates. Dense Ising coupling maps are implemented directly without SWAP overhead. The QPUs use acousto-
optic deflector-based optical control for independent beam
steering to individual ions, reducing alignment errors ~\cite{Kim:2008ApOpt,Pogorelov:2021PRXQ}, together with automated calibration software for scalable
operation. Measurement debiasing follows the batched protocol of Maksymov et al.~\cite{maksymov2023}: shots are grouped in batches of 25 and averaged, with leakage checks on the Barium system.

\subsection{Post-Processing heuristics}

The goal of post-processing in this context is to correct unfeasible sequences, e.g. sequences with overlapping beads, or sequences whose contact qubits do not match the geometry ones. Two post-processing pipelines are applied identically to BF-DCQO and the random samples. The first one is a consensus pipeline, which selects the top-$k$ samples by raw Ising energy ($k = 2,000$), computes the per-contact-qubit expectation values $\langle\sigma_j^z\rangle$ and round to a consensus contact bitstring. This can be seen either as taking the majority vote of the contact qubits configuration, or computing the signed bias fields from the contact sector. Then, it scores all feasible geometries, from a pool of $200$ candidates, against the consensus using the full Hamiltonian. Next, it selects the best geometry-contact combination and perform a final repair of the contact qubits. The second method is a per-sample repair, which filters for geometric feasibility, i.e. no backtracking. Then, it re-optimizes contact qubits independently for each feasible geometry by exhaustive enumeration of the contact Hamiltonian. Additionally, it performs a problem-inspired greedy local descent on the full bitstring. This pipeline processes each sample independently and re-optimizes contacts from scratch at each step.

The consensus pipeline aggregates information across many BF-DCQO samples, extracting a robust contact signal before combining it with geometry candidates. Per-sample repair treats each sample in isolation and can overwrite the contact structure learned by the quantum algorithm. These two post-processing techniques exploit the structure of the problem and can be considered as independent heuristic baselines, which can either start from a random bitstring or be warm-started from a quantum algorithm. 

\section{Results and discussion}
\label{sec:results}

\begin{figure}[t]
\centering
\includegraphics[width=\columnwidth]{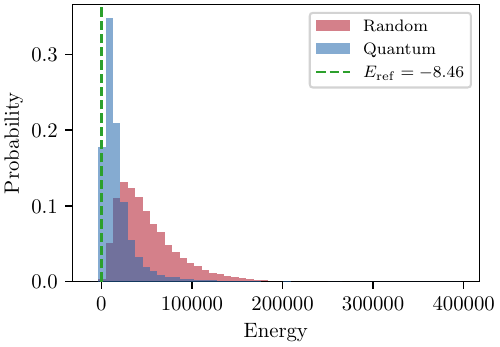}
\caption{Raw energy distributions for LEPFSGKALCSWSIC (53~qubits, high pruning). BF-DCQO (blue) produces samples with a mean energy $2.9\times$ lower than the random case (red). The dashed line marks $E_{\text{ref}} = -8.46$.}
\label{fig:raw}
\end{figure}

\begin{figure}[t]
\centering
\includegraphics[width=\columnwidth]{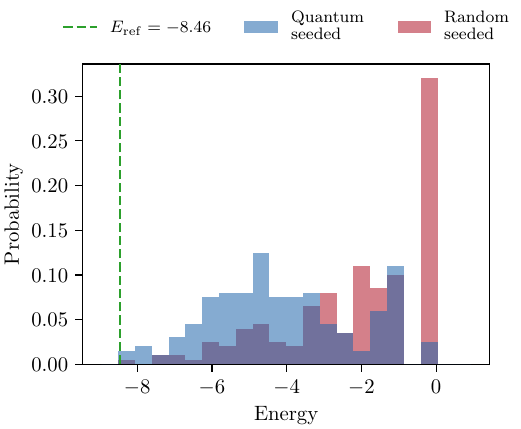}
\caption{Consensus pipeline energy distributions for LEPFSGKALCSWSIC ($53$~qubits, high pruning). Using quantum seeds (blue) is clearly shifted to lower energies compared to using random seeds (red).}
\label{fig:consensus}
\end{figure}

\begin{figure*}[t]
\centering
\includegraphics[width=\textwidth]{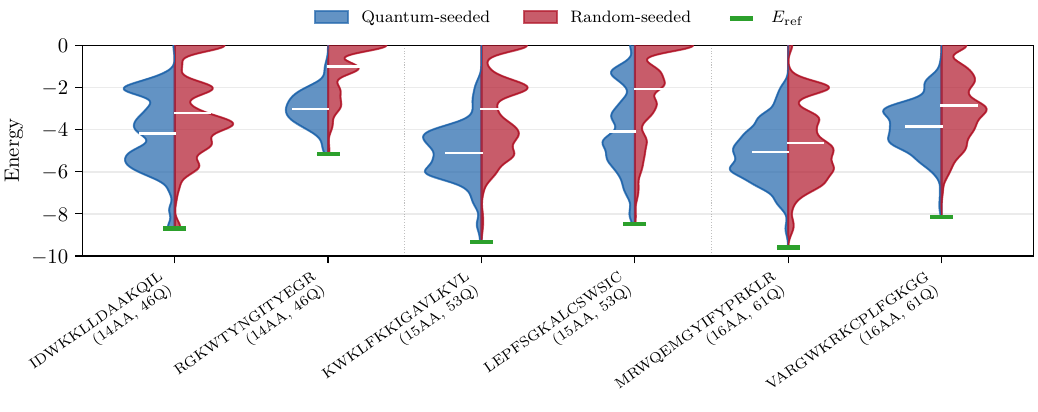}
\caption{Split-violin plot of consensus pipeline energy distributions across all 6 sequences (best pruning per sequence). Left half (blue): quantum (BF-DCQO); right half (red): random baseline. White lines indicate mean energy. Green lines show $E_{\text{ref}}$. 
}
\label{fig:summary}
\end{figure*}

\begin{figure}[h]
\centering
\includegraphics[width=0.9\linewidth]{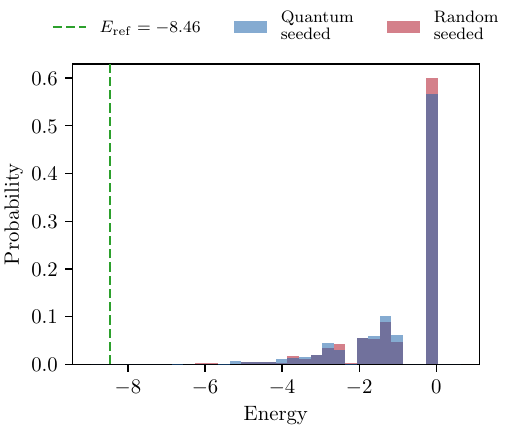}
\caption{Per-sample repair energy distributions for LEPFSGKALCSWSIC ($53$~qubits, high pruning). Using quantum seeds (blue) yields comparable energies to using random seeds (red).
}
\label{fig:pipeline_a}
\end{figure}

\subsection{Learning the Interaction Structure}

In this section, we take the sequence LEPFSGKALCSWSIC as an example. The raw energy distributions provide direct evidence that BF-DCQO learns meaningful problem structure. The reference energies $E_\mathrm{ref}$ are obtained by running a classical genetic algorithm until convergence. Specifically, Fig.~\ref{fig:raw} shows that BF-DCQO produces samples with average raw energy $1.8 \times 10^4$, compared to $5.2 \times 10^4$ for the random baseline, a $2.9\times$ reduction. This means that the results were not completely degraded by hardware noise. Nevertheless, the raw energies are still large and positive because the majority of bitstrings violate geometric constraints. Only ${\sim}3.1\%$ of the samples are geometrically feasible. There are many mechanisms for this to happen, on the algorithmic side, the quantum circuits do not preserve the constraints natively, so that unfeasible bitstrings are still expected to be populated even under ideal conditions; and on the hardware side, it is sufficient to have one single spin-flip to completely change the geometry of the sequence and make it infeasible. However, the key observation is that BF-DCQO's distribution is consistently shifted toward lower energies, indicating that the iterative bias updates work as expected at this scale. The main source of having lower energy samples comes from the contact qubits, where the per-qubit projections from BF-DCQO's top samples show strong polarization away from $0.5$, mostly concentrated near $0$ or $1$, compared to the random case ${\sim}0.45$.

\subsection{Applying the Consensus Heuristic}

The consensus pipeline preserves BF-DCQO's learned contact patterns. Figure~\ref{fig:consensus} shows the energy distributions after applying the post-processing heuristic for the LEPFSGKALCSWSIC sequence. The distribution is clearly shifted toward lower energies when quantum states are used as seeds, yielding a mean of $-4.09$ vs. $-2.07$ for the random seeds. Additionally, in the quantum case, the reference energy $E_{\text{ref}} = -8.46$ is obtained, unlike in the randomly seeded case. Across all instances, the warm-started consensus pipeline reaches the classical reference energy in $4$ out of $6$ sequences. In the randomly-started case, it reaches the reference energy in $1$ out of $6$ sequences.

Figure~\ref{fig:summary} provides a visual summary of the consensus pipeline across all sequences. At $46$ and $53$ qubits, using BF-DCQO shows a consistent advantage, reaching the reference energy in all four sequences at these sizes. At $61$ qubits, the advantage diminishes as the $\theta$ parameter in Algorithm~\ref{alg:bfdcqo} was chosen to keep the number of entangling gates in the range of $1,000$, which affected the quality of the result at this scale. As a consequence, none of the two sequences reached the reference energy via the consensus pipeline.

\subsection{Effect of Post-Processing}

\begin{table*}[htb]
\centering
\small
\setlength{\tabcolsep}{4pt}

\caption{Energy results across all peptide sequences for the two post-processing heuristics. $E^\mathrm{Q}$ corresponds to obtained energies after using quantum seeds, whereas $E^\mathrm{R}$ refers to random seeds.}
\label{tab:results}

\begin{tabularx}{\textwidth}{>{\raggedright\arraybackslash}X r r r r r r r r r}
\toprule
& & \multicolumn{4}{c}{Consensus} & \multicolumn{4}{c}{Per-sample repair} \\
\cmidrule(lr){3-6} \cmidrule(lr){7-10}
Sequence & $E_{\mathrm{ref}}$
& $E_\mathrm{avg}^{Q}$ & $E_\mathrm{avg}^{R}$ & $E_\mathrm{best}^{Q}$ & $E_\mathrm{best}^{R}$
& $E_\mathrm{avg}^{Q}$ & $E_\mathrm{avg}^{R}$ & $E_\mathrm{best}^{Q}$ & $E_\mathrm{best}^{R}$ \\
\midrule
IDWKKLLDAAKQIL
& $-8.698$
& $-4.185$ & $-3.206$ & $\mathbf{-8.698}$ & $\mathbf{-8.698}$
& $-2.595$ & $-2.556$ & $\mathbf{-8.698}$ & $\mathbf{-8.698}$ \\

RGKWTYNGITYEGR
& $-5.162$
& $-3.034$ & $-1.013$ & $\mathbf{-5.162}$ & $-5.042$
& $-1.807$ & $-1.772$ & $-5.089$ & $-5.042$ \\

KWKLFKKIGAVLKVL
& $-9.321$
& $-5.109$ & $-3.032$ & $\mathbf{-9.321}$ & $-8.629$
& $-3.450$ & $-3.304$ & $\mathbf{-9.321}$ & $\mathbf{-9.321}$ \\

LEPFSGKALCSWSIC
& $-8.463$
& $-4.085$ & $-2.065$ & $\mathbf{-8.463}$ & $-8.147$
& $-3.113$ & $-3.042$ & $-8.286$ & $-8.159$ \\

MRWQEMGYIFYPRKLR
& $-9.598$
& $-5.074$ & $-4.649$ & $-8.767$ & $-8.890$
& $-3.560$ & $-3.338$ & $-8.819$ & $-9.010$ \\

VARGWKRKCPLFGKGG
& $-8.155$
& $-3.847$ & $-2.851$ & $-7.793$ & $-6.925$
& $-2.653$ & $-2.535$ & $-7.179$ & $-7.450$ \\
\bottomrule
\end{tabularx}
\end{table*}

While the consensus heuristic is benefited from the quantum seeds, there exist heuristics that effectively overwrite the learned pattern from the quantum algorithm, as discussed in Sec.~\ref{sec:results}A. One such example is the per-sample repair. Figure~\ref{fig:pipeline_a} illustrates this for the LEPFSGKALCSWSIC instance across three stages: raw samples, contact-repaired, and after greedy descent. BF-DCQO seeds start with a clear advantage in raw energy, which diminishes as geometries are corrected and the contacts are re-optimized, hence overwritten. This leads to an underestimation of the quantum algorithm's contribution, as this heuristic abruptly erases the contact qubits, which correspond to the majority of the total number of qubits. The consensus pipeline avoids this by preserving the aggregate contacts from the quantum samples.

In Eq.~\ref{eq:hamiltonian}, the contact qubits are effectively unconstrained binary variables, given any backbone geometry, their optimal values are determined by the interaction Hamiltonian. The quantum algorithm excelled on these degrees of freedom. Despite having a good estimation of the contribution from the contact qubits, the challenge of finding the optimal backbone geometry remains. While it is trivial to estimate the right contact qubits from a fixed configuration of the geometry qubits, the converse is not. Hence, we relied on post-processing classical heuristics to do so.

Our comparison of per-sample repair and the consensus pipeline highlights that classical post-processing must be designed to exploit the quantum samples. The exact values are presented in Table~\ref{tab:results}. Naively choosing the post-processing, as exemplified in Fig.~\ref{fig:pipeline_a}, can lead to overwriting the preliminary advantage from the quantum seeds.

\section{Conclusion}

We have demonstrated bias-field digitized counterdiabatic quantum optimization (BF-DCQO) for lattice protein folding on a 64-qubit Barium development system similar to the forthcoming IonQ Tempo line, studying six peptide instances requiring 46--61 qubits. To the best of our knowledge, this is the largest trapped-ion demonstration of quantum protein-folding optimization to date, and the largest realization on digital quantum hardware using the full dense tetrahedral encoding, originally introduced in Ref.~\cite{robert2021}

The main result is not simply that BF-DCQO lowers the raw energy relative to uniform random sampling, but that the improvement has a clear structure. However, because the quantum circuits do not enforce geometric feasibility and the presence of hardware noise, most raw samples still violate backbone constraints. The advantage is instead observed in the interaction sector of the Hamiltonian, which constitute the majority of the qubits. This behavior is consistent with the bias-field update mechanism, where low-energy samples polarize the local fields, allowing subsequent rounds to reinforce favorable interaction patterns. 

The comparison between the two post-processing strategies highlights an important point for hybrid quantum-classical optimization for protein folding. A repair routine that treats each sample independently and re-optimizes the contact variables can remove part of the information generated by the quantum circuit. In contrast, the consensus-based pipeline aggregates contact information across many low-energy BF-DCQO samples before combining it with feasible backbone geometries. This preserves the quantum-learned contact structure and leads to improved energy distributions compared with the corresponding random-seeded workflow. For constrained problems, such as lattice-based protein folding, near-term quantum processors may be most useful when they provide structured statistical information. In this study, BF-DCQO supplies such information in the residue-contact sector, while classical post-processing enforces geometric feasibility and assembles complete candidate folds. This division of labor provides a practical route for using current quantum hardware on structured optimization problems with strong constraints.

Future improvements should aim to make the quantum part of the workflow more aware of the protein-folding structure. Constraint-preserving drivers could reduce the number of invalid backbone configurations, while separate bias-field treatments for geometry and contact qubits may better reflect their different roles in the encoding. Higher-order counterdiabatic ansatzes and decomposition strategies could also help capture correlations that are not accessible with the present circuit structure. Together, these directions could strengthen the link between quantum-learned contact patterns and feasible low-energy protein conformations, enabling larger and more accurate trapped-ion demonstrations of quantum protein folding.

\bibliographystyle{IEEEtran}
\bibliography{references}

\end{document}